\documentclass{sig-alternate}

\usepackage{color} 
\graphicspath{{graphics/}}
\usepackage{url}

\newcommand{\x}{x}

\newcommand{\kth}[1]{k_{#1}}

\newcommand{\M}{\textnormal{M}}
\newcommand{\bigS}{\textnormal{S}}
\newcommand{\A}{A}
\newcommand{\Dx}[1]{D_{#1}}
\newcommand{\Nx}[1]{N_{#1}}
\newcommand{\metre}{\textnormal{m}}
\newcommand{\second}{\textnormal{s}}

\newcommand{\EXP}[1]{\exp\left(#1\right)}

\newcommand{\Vol}[1]{V_{#1}}

\newcommand{\prop}[1]{a_{#1}}
\newcommand{\molNum}[1]{U_{#1}}

\newfont{\mycrnotice}{ptmr8t at 7pt}
\newfont{\myconfname}{ptmri8t at 7pt}
\let\crnotice\mycrnotice%
\let\confname\myconfname%

\begin{document}\sloppy
%
\permission{Permission to make digital or hard copies of all or part of this work for personal or classroom use is granted without fee provided that copies are not made or distributed for profit or commercial advantage and that copies bear this notice and the full citation on the first page. Copyrights for components of this work owned by others than the author(s) 
must be honored. Abstracting with credit is permitted. To copy otherwise, or republish, to post on servers or to redistribute to lists, requires prior specific permission and/or a fee. Request permissions from Permissions@acm.org.}
\conferenceinfo{NANOCOM '15,}{September 21 -- 22 2015, Boston, MA, USA\\
{\mycrnotice{Copyright is held by the owner/author(s). Publication rights 
licensed to ACM.}}}
\CopyrightYear{2015} 
\copyrightetc{ACM \the\acmcopyr}

\title{On the Statistics of Reaction-Diffusion Simulations for Molecular Communication}

%
%
%
%
%

\numberofauthors{3} 
%
\author{
%
%
\alignauthor
Adam Noel\\
       \affaddr{Department of Electrical and Computer Engineering}\\
       \affaddr{University of British Columbia}\\
       \affaddr{Vancouver, BC, Canada}\\
       \email{adamn@ece.ubc.ca}
\alignauthor
Karen C. Cheung\\
       \affaddr{Department of Electrical and Computer Engineering}\\
       \affaddr{University of British Columbia}\\
       \affaddr{Vancouver, BC, Canada}\\
       \email{kcheung@ece.ubc.ca}
\alignauthor Robert Schober\\
       \affaddr{Institute for Digital Communications}\\
       \affaddr{Friedrich-Alexander-Universit\"{a}t Erlangen-N\"{u}rnberg}\\
       \affaddr{Erlangen, Germany}\\
       \email{schober@lnt.de}
}
\date{16 May 2015}

\maketitle

\begin{abstract}	
	A molecule traveling in a realistic propagation environment can experience stochastic interactions with other molecules and the environment boundary. The statistical behavior of some isolated phenomena, such as dilute unbounded molecular diffusion, are well understood. However, the coupling of multiple interactions can impede closed-form analysis, such that simulations are required to determine the statistics. This paper compares the statistics of molecular reaction-diffusion simulation models from the perspective of molecular communication systems. Microscopic methods track the location and state of every molecule, whereas mesoscopic methods partition the environment into virtual containers that hold molecules. The properties of each model are described and compared with a hybrid of both models. Simulation results also assess the accuracy of Poisson and Gaussian approximations of the underlying Binomial statistics.
\end{abstract}



\section{Introduction}

The prevalence of using molecules to communicate in biological systems (see \cite[Ch.~16]{RefWorks:588}) has recently attracted the attention of the research community to adapt the principles of molecular communication (MC) for new applications to transmit arbitrary amounts of information in environments where conventional methods of communication might be hazardous or impractical; see \cite{RefWorks:801}. One MC method, free diffusion, is attractive because it does not require additional infrastructure in the propagation medium. Free diffusion is effectively a random process where a molecule collides with other molecules in a fluid environment.

The behavior of any one molecule in a realistic propagation environment is unlikely to be characterized by diffusion alone. Other potential phenomena include bulk fluid flow, collisions with the environment boundary, and chemical reactions either throughout the environment or in a local region. Generally, these phenomena contribute to the stochastic behavior of any single molecule.

We noted in \cite{RefWorks:891} that communications analysis requires the form and the statistics of the end-to-end channel impulse response, i.e., the time-varying signal observed at the receiver given that molecules are released at some instant by the transmitter. The response can then be used to derive the received signal for any modulation scheme. Analytical models for some isolated processes are known, such as for molecular diffusion; see \cite{RefWorks:586}. However, when multiple interactions are present, their impact is coupled and this can impede closed-form theoretical analysis. Often, simplifying assumptions are made and specific geometries are studied to facilitate analysis. For example, we analyzed an unbounded environment with diffusion, bulk fluid flow, and molecule degradation in \cite{RefWorks:747}. A closed-form time domain channel impulse response was derived, but this was in the absence of any local chemical reactions (such as at the receiver). Generally, we may need to rely on numerical methods or simulations to determine the channel statistics.

Simulation methods for molecular behavior can range in scale from \emph{molecular dynamics} models (such as that used in LAMMPS \cite{RefWorks:943}), which account for all interactions between all individual molecules (including solvent molecules in a fluid), to \emph{continuum} models (such as that used in COMSOL Multiphysics \cite{RefWorks:944}) where no individual molecules are described. Two common ``intermediate'' models that tend to be suitable for the study of reaction-diffusion environments are \emph{microscopic} and \emph{mesoscopic} models. Both of these models treat the solvent in a fluid as a continuum and focus on the behavior of solute molecules.

Microscopic simulators such as the Smoldyn simulator track the coordinates and behavior of each solute molecule; see \cite{RefWorks:622}. Mesoscopic simulators partition the system into virtual containers and track the number and type of solute molecules in each container. If molecular concentrations in each container are homogeneous, then a mesoscopic simulation can accurately capture the behavior of the system; see \cite{RefWorks:617}. However, the assumption of homogeneity can place severe constraints on the size of virtual containers; see \cite{RefWorks:613,RefWorks:617}. A microscopic model has better spatial accuracy, but the advantages of a mesoscopic model include easier implementation of complex chemical reactions and better computational efficiency as the system dimensions grow.

From a MC perspective, we are ultimately interested in the accuracy of the statistics at the receiver. We may need to simulate the system many thousands of times to compile the receiver statistics. The behavior of the total system is not as important, so we are motivated to improve computational efficiency in regions that  are not critical to the receiver statistics, i.e., are sufficiently far from the communication link. In \cite{RefWorks:891}, we combined two schemes towards this goal. In the first scheme, also described in \cite{RefWorks:806}, the local accuracy is adjusted by using mesoscopic containers of different sizes; generally, larger subvolumes are less accurate but more computationally efficient. In the second scheme, the environment is partitioned into microscopic and mesoscopic regimes, thus providing additional flexibility in the tradeoff between local accuracy and computational complexity. These hybrid schemes have been proposed in papers including \cite{RefWorks:870,RefWorks:851,RefWorks:872}.

In this paper, we use simulations to study the accuracy of both the average time-varying channel response and the statistics of the response at specific times. Unlike in \cite{RefWorks:891}, where we observed the channel response due to diffusion only, here we study the channel response and statistics of diffusion, a first-order chemical reaction, and a simple reaction-diffusion scenario. We gain insight into how aggressively we can reduce the computational complexity of the simulation environment without compromising the accuracy at the receiver. A formal study of computational complexity is left for future work, but preliminary results were shown in \cite{RefWorks:891}.

We also compare the suitability of the Poisson and Gaussian approximations of the Binomial distribution when representing the cumulative distribution function (CDF) of receiver observations made at a specific time. These approximations are commonly used in communications analysis because they are less computationally intensive, and were also recently assessed for MC systems in \cite{RefWorks:910}.

The rest of this paper is organized as follows. The underlying physical model and channel statistics are described in Section~\ref{sec_phys_model}. The microscopic, mesoscopic, and hybrid simulation models are briefly defined in Section~\ref{sec_sim_model}. Simulation results to compare the models are presented in Section~\ref{sec_results}. Section~\ref{sec_concl}
concludes the paper.

\section{Physical Model}
\label{sec_phys_model}

In this section, we describe the common physical model that the simulation models represent. We present the expected channel responses for the scenarios that we will simulate, and discuss the statistics of those responses.

The environment is a bounded two-dimensional fluid ``volume'' $\Vol{}$ with a reflective boundary; generally, it could be absorbing or reactive if there are local chemical reactions at the boundary. There is a single molecular species, labeled molecule $\A$, with constant diffusion coefficient $\Dx{}$. Using a constant $\Dx{}$ implies that the $\A$ molecules are dilute.

We consider two diffusive scenarios where the expected channel response at the receiver (RX) and the corresponding statistics are (at least approximately) known so that we can focus on assessing the accuracy of the simulation models. In the first scenario, we distribute molecules uniformly over $\Vol{}$ and observe the number of molecules present in $\Vol{RX}$, a subset of $\Vol{}$. If $\Nx{}$ $\A$ molecules are distributed, then the number of molecules \emph{expected} in $\Vol{RX}$, $\overline{\molNum{RX}}(t)$, is constant and equal to
\begin{equation}
\label{eq_mol_exp_uni}
\overline{\molNum{RX}}(t) =\Nx{}\Vol{RX}/\Vol{}.
\end{equation}

In the second scenario, we release (i.e., transmit) $\Nx{}$ $\A$ molecules in a small area near the center of $\Vol{}$ and observe (i.e., receive) the number of molecules present in another small area also near the center of $\Vol{}$. If $\Vol{}$ is large enough to model as infinite, and the transmitter (TX) and RX regions are small enough to model as points, then from \cite[Eq.~(3.4)]{RefWorks:586} we can write  $\overline{\molNum{RX}}(t)$ as
\begin{equation}
\label{eq_mol_exp_p2p}
\overline{\molNum{RX}}(t) = \frac{Nh_{RX}^2}{4\pi\Dx{} t}\EXP{-\frac{d^2}{4\Dx{}t}},
\end{equation}
where $h_{RX}^2$ is the area of the RX, and $d$ is the distance between the centers of the TX and the RX. Eq.~(\ref{eq_mol_exp_p2p}) is accurate if $h_{RX} \ll d$ and $d^2 \ll \Vol{}$.

An $\A$ molecule can also degrade according to the first-order chemical reaction $\A \xrightarrow{\kth{}} \emptyset$, where $\kth{}$ is the reaction rate constant in $\second^{-1}$. If there are $\Nx{}$ $\A$ molecules in the system at time $t=0\,\second$, and the RX is \emph{all} of $\Vol{}$, then the number of molecules \emph{expected} to remain at time $t>0$, is \cite[Eq.~(9.7)]{RefWorks:585}
\begin{equation}
\label{eq_mol_exp_rxn}
\overline{\molNum{RX}}(t) = \Nx{}\EXP{-\kth{}t},
\end{equation}
such that each molecule has a probability of $\EXP{-\kth{}t}$ of remaining at time $t$, and we can account for degradation in the diffusive scenarios by scaling (\ref{eq_mol_exp_uni}) or (\ref{eq_mol_exp_p2p}) by $\EXP{-\kth{}t}$.

Now consider the statistics of the channel responses at a specific instant $t$. Assuming no knowledge of one molecule's location or whether it has been degraded after time $t>0$, then whether that molecule is in the RX at time $t$ is the outcome of an independent trial; see \cite[Ch.~5.1]{RefWorks:725}. For general $\Nx{}$, there is one trial for each molecule. The number of molecules observed in the RX at time $t$, $\molNum{RX}(t)$, is the number of ``successful'' trials. Thus, $\molNum{RX}(t)$ for some $t$ is a Binomial random variable, where the probability of success of each trial is equal to $\overline{\molNum{RX}}(t)$ with $\Nx{}=1$.

By knowing $\overline{\molNum{RX}}(t)$, we can compare the empirical CDF of each simulation model with the Binomial CDF. We can also assess the Poisson and Gaussian approximations of the Binomial CDF. From \cite[Ch.~5.2]{RefWorks:725}, the Poisson approximation should be accurate when $\Nx{}$ is ``large'' and $\overline{\molNum{RX}}(t)$ for $\Nx{}=1$ is ``small''. From \cite[Ch.~5.5]{RefWorks:725} and the Central Limit Theorem, the Gaussian approximation should be accurate for sufficiently large $\Nx{}$.

\section{Simulation Models}
\label{sec_sim_model}

In this section, we summarize the simulation models that we will assess in Section~\ref{sec_results}.

\subsection{Microscopic Model}
\label{sec_micro}

In the microscopic model, the environment $\Vol{}$ is a single container $\Vol{\M}$ and there is a constant time step $\Delta t_{\M}$. For each time step, the coordinates of every $\A$ molecule are updated by adding a random displacement $n\sqrt{2\Dx{}\Delta t_{\M}}$ to each dimension, where $n$ is an independent normal random value with mean 0 and variance 1. Any molecule that ends up outside of $\Vol{\M}$ is reflected against the boundary of $\Vol{\M}$. A given molecule is degraded during the time step and removed from  $\Vol{}$ if $u > \EXP{-\kth{}\Delta t_{\M}}$, where $u$ is an independent uniform random value between 0 and 1.

\subsection{Mesoscopic Model}
\label{sec_meso}

In the mesoscopic model, $\Vol{}$ is partitioned into virtual subvolumes or containers. We track the number of $\A$ molecules within each subvolume and assume that each subvolume is homogeneous. Mesoscopic simulations are described as a series of ``events''. A diffusion event is the transition of a molecule between adjacent subvolumes, and a degradation event is a decrement of the number of molecules in one subvolume. Every event is assigned a \emph{propensity}, $\prop{}$, which determines the probability of that event occurring next. In this paper, we summarize how the relevant propensities are calculated. Due to space, the reader is referred to \cite{RefWorks:891} or to works such as \cite{RefWorks:613} for details on how to use the propensities to simulate an event sequence.

Diffusion propensities describe the expected transition rate of molecules between adjacent subvolumes. Following \cite{RefWorks:891}, we consider square subvolumes that could have different sizes in order to adjust the local computational complexity. The transition rate from a square subvolume of width $h_i$ to one of width $h_j$, where the overlap of their adjacent faces is length $h_o \le \min\{h_i,h_j\}$, is found to be \cite[Eq.~(9)]{RefWorks:891}
\begin{equation}
\label{eq_diff_rate}
\prop{i,j} = \frac{2\Dx{}h_o\molNum{i}}{h_i^2(h_i + h_j)},
\end{equation}
where $\molNum{i}$ is the number of molecules currently in subvolume $i$. The propensity of a chemical reaction describes the expected frequency of the reaction. For our first-order degradation, the propensity $\prop{\textnormal{rxn},i}$ of the reaction in subvolume $i$ is given by \cite[Eq.~(6)]{RefWorks:613} as $\prop{\textnormal{rxn},i} = \kth{}\molNum{i}$.

\subsection{Hybrid Model}
\label{sec_hybrid}

In the hybrid model, $\Vol{}$ is partitioned into a microscopic regime $\Vol{\M}$ and a mesoscopic regime $\Vol{\textnormal{S}}$. Both regimes are treated independently, as previously described, until there are molecules that transition from one regime to the other. For simplicity, as in \cite{RefWorks:891}, we adopt the simplified transition rules described in \cite{RefWorks:870}:

\emph{$\Vol{\bigS}$ to $\Vol{\M}$}: A source subvolume in $\Vol{\bigS}$ must be along the boundary with $\Vol{\M}$, and it has a mirror ``imaginary'' subvolume of the same size in $\Vol{\M}$. A molecule leaving $\Vol{\bigS}$ to enter $\Vol{\M}$ is placed at random within the mirror subvolume and then treated as an individual molecule in $\Vol{\M}$.

\emph{$\Vol{\M}$ to $\Vol{\bigS}$}: When a molecule in $\Vol{\M}$ is identified to have entered $\Vol{\bigS}$, then we add that molecule to the subvolume along the boundary with $\Vol{\M}$ that is closest to the molecule's new location.

\section{Numerical Results}
\label{sec_results}

\begin{figure}[!tb]
	\centering
	\def\svgwidth{\linewidth}
	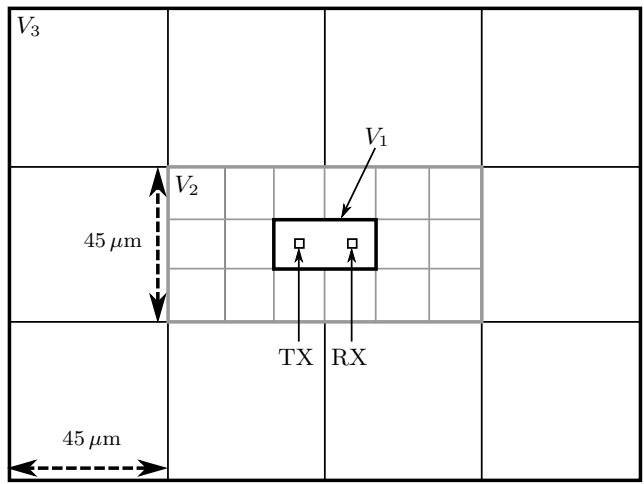
	\caption{Simulation environment $\Vol{}$ drawn to scale. The inner region $\Vol{1}$ has width $30\,\mu$m and height $15\,\mu$m. The TX and RX are squares of width $3\,\mu$m, are placed in the middle of $\Vol{1}$, and are separated by a distance of $15\,\mu$m (center to center). The middle region $\Vol{2}$ (in grey) surrounds $\Vol{1}$ and has width $90\,\mu$m. The outer region $\Vol{3}$ (in black) surrounds $\Vol{2}$, has width $180\,\mu$m, and has a reflective outer boundary. The partitioning of the regions of $\Vol{}$ shown here is an example of the HYB model in Table~\ref{table_partition}.}
	\label{fig_sim_model}
\end{figure}

In this section, we present simulation results to assess the accuracy of the simulation models to generate the channel response and the channel statistics. We consider the environment defined in Fig.~\ref{fig_sim_model} (or a subset of it) for all of our simulations. The coefficient of diffusion is $\Dx{}=10^{-9}\frac{\metre^2}{\second}$. The environment $\Vol{}$ is partitioned into 3 regions. Each region is modeled as either microscopic or mesoscopic. Unless otherwise specified, the model labels used in the simulation figures are described in Table~\ref{table_partition}. Namely, we consider microscopic (MICRO), mesoscopic (MESO), multi-scale MESO (MESO-MS) and hybrid (HYB) partitioning models, where both the MESO-MS and HYB models are less accurate in $\Vol{2}$ and/or $\Vol{3}$ because the communication link is in $\Vol{1}$.

\begin{table}[!tb]
	\centering
	\caption{System partitioning models, unless otherwise specified. When a region is mesoscopic, all subvolumes in that region have the specified width. $\Vol{2}$ for a HYB model is mesoscopic if $h_2$ is defined.}
	{\renewcommand{\arraystretch}{1.4}
		\begin{tabular}{|c|c|c|c|c|}
			\hline
			\bfseries Model & $\Vol{1}$ & $\Vol{2}$ & $\Vol{3}$ \\ \hline
			MICRO & Micro & Micro & Micro \\ \hline
			MESO & $h$ & $h$ & $h$ \\ \hline
			MESO-MS & $h_1$ & $h_2$ & $h_3$ \\ \hline
			HYB & Micro & Micro or $h_2$ & $h_3$ \\ \hline
		\end{tabular}
	}
	\label{table_partition}
\end{table}

\begin{figure}[!tb]
	\centering
	\includegraphics[width=\linewidth]{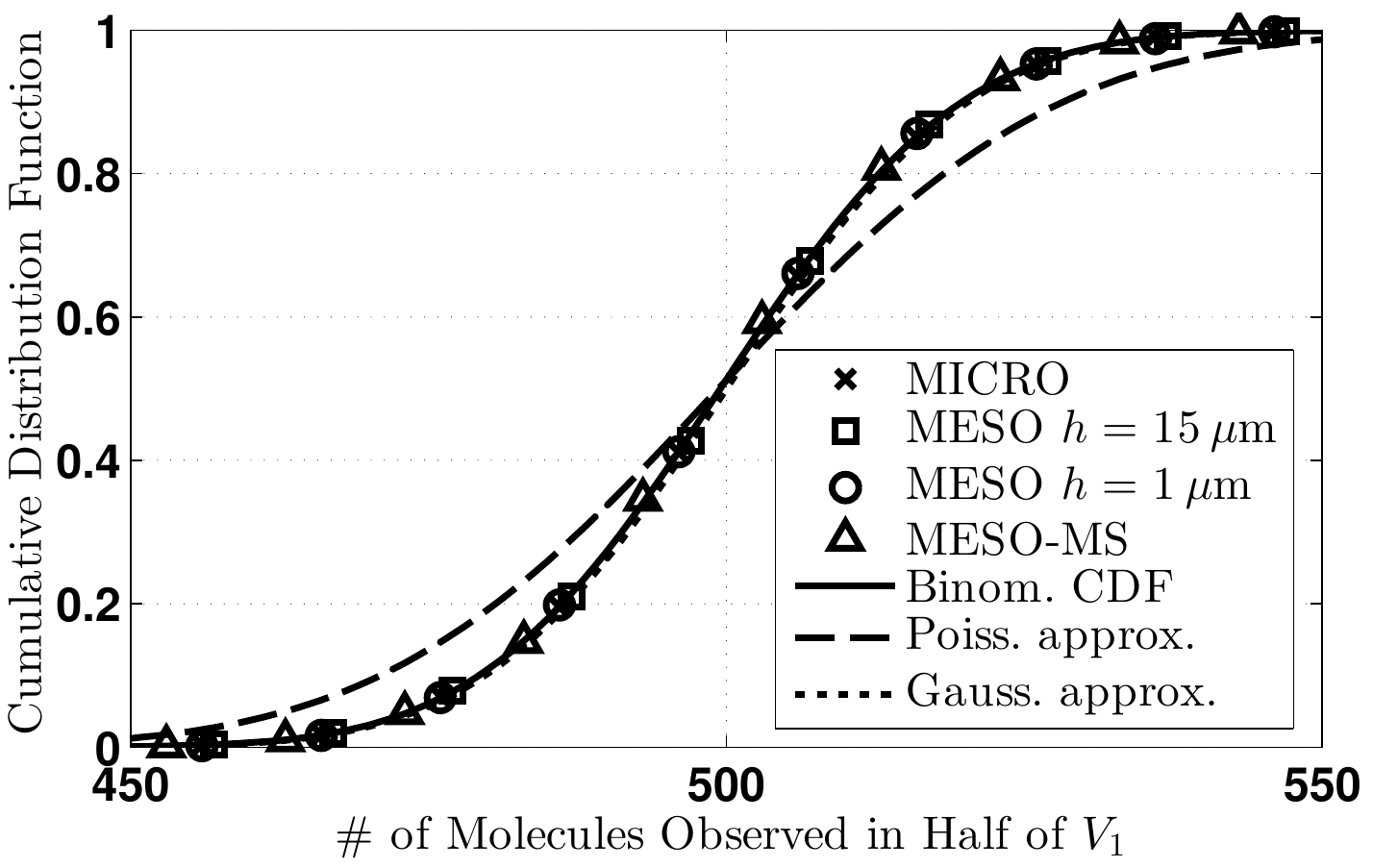}
	\caption{CDF of simulation 1, i.e., the uniform diffusion test where molecules are restricted to $\Vol{1}$ \emph{only}. The observation is made at time $t=5\,$s.}
	\label{fig_sim_04_uni_cdf}
\end{figure}

We describe a series of 5 simulations as summarized in Table~\ref{table_test_summary}. The first two simulations are uniform diffusion tests. For the first test, the system consists of region $\Vol{1}$ \emph{only} (this is the only test where molecule motion is restricted to $\Vol{1}$). 1000 molecules are initialized over all of $\Vol{1}$ and we observe the number of molecules present in one half of the region after $t=5\,\second$, i.e., 500 molecules are expected. The empirical CDF for each partitioning model, compiled over $10^4$ realizations, is presented in Fig.~\ref{fig_sim_04_uni_cdf}. The empirical CDFs of the simulation models all match the Binomial CDF, including the MESO-MS model where here we observe the number of molecules in a subvolume with width $15\,\mu$m and the rest of region is partitioned into subvolumes with width $1\,\mu$m. The Gaussian approximation is very close to the Binomial CDF whereas the Poisson approximation is not, since the underlying trial success probability (i.e., the probability that a given molecule is observed) is 0.5, which is very high.

\begin{table}[!tb]
	\centering
	\caption{Simulation test parameter summary. The units for $\kth{}$ and $\Delta t_{\M}$ are $\second^{-1}$ and ms, respectively.}
	{\renewcommand{\arraystretch}{1.3}
		\begin{tabular}{|c|c|c|c|c|c|c|}
			\hline
			\bfseries Test & \bfseries System & \bfseries Source & \bfseries Observer & $\Nx{}$ & $\kth{}$ & $\Delta t_{\M}$ \\ \hline
			1 & $\Vol{1}$ & $\Vol{1}$ & Half of $\Vol{1}$ & 1000 & 0 & 5 \\ \hline
			2 & All & All & $\Vol{1}$ & 10800 & 0 & 10 \\ \hline
			3 & All & TX & RX & 1500 & 0 & 5 \\ \hline
			4 & - & - & - & 1000 & 3 & Vary \\ \hline
			5 & All & TX & RX & 1500 & 3 & 5 \\ \hline
		\end{tabular}
	}
	\label{table_test_summary}
\end{table}

\begin{figure}[!tb]
	\centering
	\includegraphics[width=\linewidth]{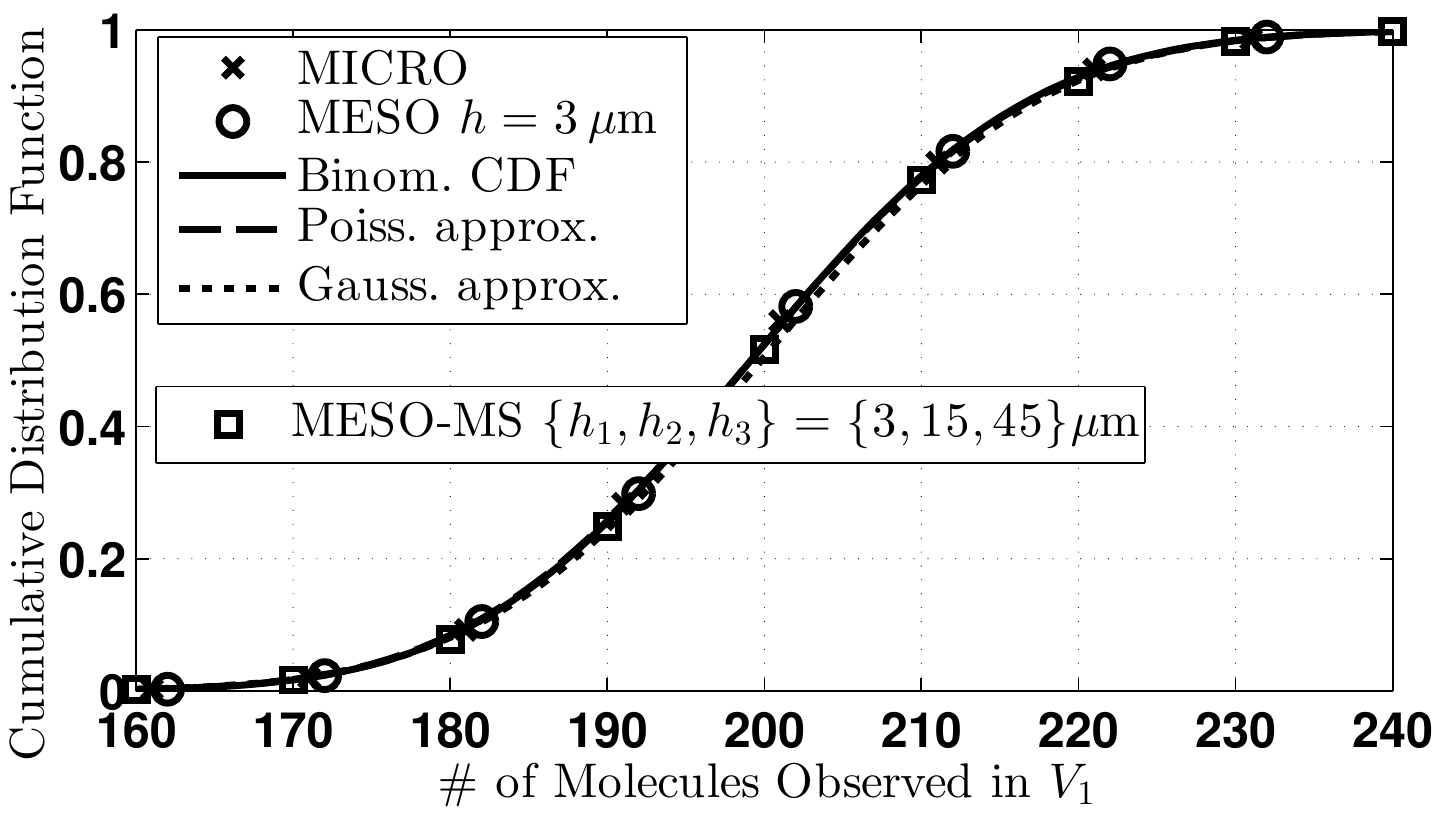}
	\caption{CDF of simulation 2, i.e., the uniform diffusion test that places molecules over all of $\Vol{}$. The observation is made at time $t=20\,$s.}
	\label{fig_sim_01_uni_cdf}
\end{figure}

In the second simulation, we perform a uniform diffusion test where we initialize 10800 molecules over all of $\Vol{}$ and we observe the number of molecules present in $\Vol{1}$ after $t=20\,\second$. From (\ref{eq_mol_exp_uni}), we expect to observe a mean of 200 molecules. The empirical CDF for each simulation model, compiled over $10^4$ realizations, is presented in Fig.~\ref{fig_sim_01_uni_cdf}. The empirical CDFs of all three models match the Binomial CDF. Here, the Poisson approximation also matches the Binomial CDF, whereas the Gaussian approximation is slightly less accurate.

\begin{figure}[!tb]
	\centering
	\includegraphics[width=\linewidth]{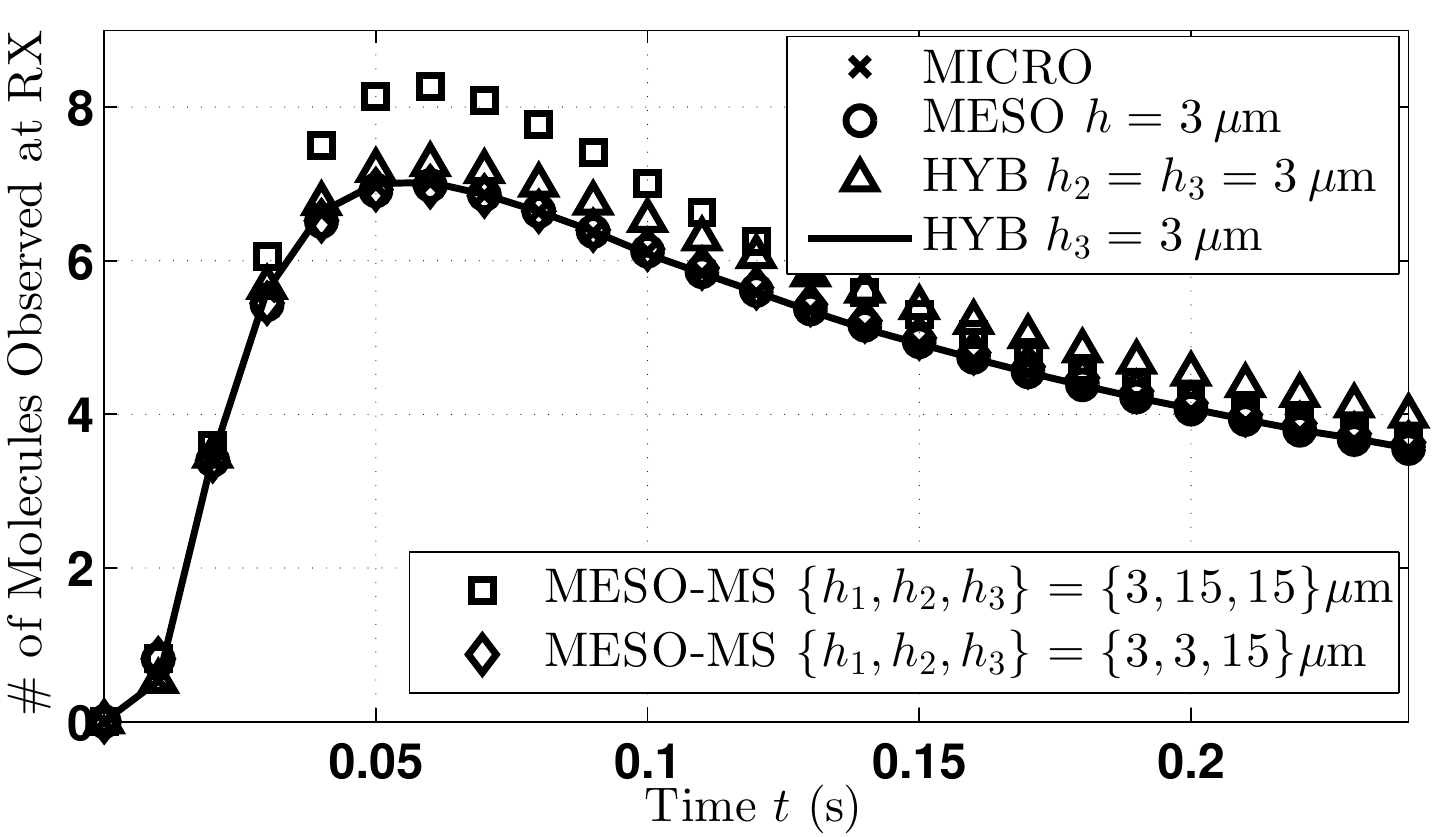}
	\caption{Time-varying channel response of simulation 3, i.e., the ``point-to-point'' diffusion test.}
	\label{fig_sim_01_p2p_avg}
\end{figure}

\begin{figure}[!tb]
	\centering
	\includegraphics[width=\linewidth]{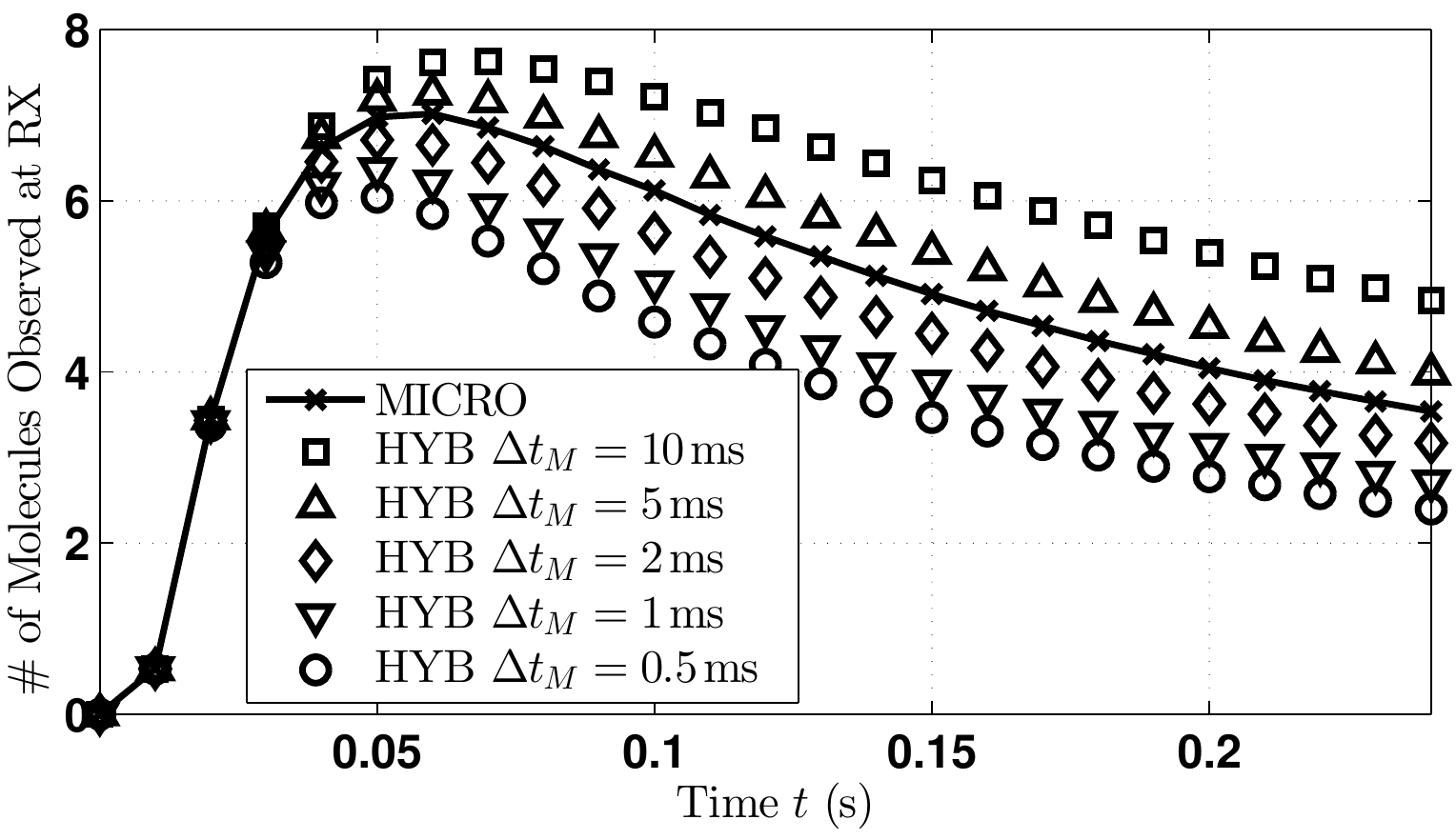}
	\caption{Time-varying channel response of simulation 3, i.e., the ``point-to-point'' diffusion test, where we highlight the impact of $\Delta t_{\M}$ on the accuracy of the HYB model where $h_2 = h_3 = 3\,\mu$m.}
	\label{fig_sim_01_p2pdt_avg}
\end{figure}

In the third simulation, we consider ``point-to-point'' diffusion, where 1500 molecules are released at the TX and then observed over time at the RX. The TX and RX are both squares of width $3\,\mu$m and are separated by a distance of $15\,\mu$m from center to center, as shown in Fig.~\ref{fig_sim_model}. The time-varying channel impulse response, averaged over $10^5$ realizations, is plotted for different simulation models in Figs.~\ref{fig_sim_01_p2p_avg} and \ref{fig_sim_01_p2pdt_avg}. We omit a curve of the expected channel impulse response, as found by evaluating (\ref{eq_mol_exp_p2p}), because it is effectively identical to that given by the MICRO simulation model.

In Fig.~\ref{fig_sim_01_p2p_avg}, we observe that the MESO model (with subvolumes of size $3\,\mu$m everywhere) and the HYB model that is microscopic in regions $\Vol{1}$ and $\Vol{2}$ are both very accurate when compared with the MICRO model. The MESO-MS model with subvolumes of size $3\,\mu$m in $\Vol{2}$ is also very accurate, but increasing those subvolumes to $15\,\mu$m leads to an 18\,\% overestimation of the channel impulse response at the time of the expected peak observation ($\sim0.55\,\second$). This excess is because, from (\ref{eq_diff_rate}), it takes longer for molecules to  diffuse to larger subvolumes, which here are too close to the TX and RX. However, this model is still asymptotically accurate over time because the transition rate (\ref{eq_diff_rate}) derived in \cite{RefWorks:891} leads to a uniform molecule distribution. Finally, the HYB model that is only microscopic in $\Vol{1}$ generally overestimates the channel impulse response (i.e., the MICRO/MESO interface is too close to the communication link). We further study the accuracy of HYB models in Fig.~\ref{fig_sim_01_p2pdt_avg}.

In Fig.~\ref{fig_sim_01_p2pdt_avg}, we compare the MICRO model with variations of the HYB model that overestimated the channel impulse response in Fig.~\ref{fig_sim_01_p2p_avg}, i.e., where regions $\Vol{2}$ and $\Vol{3}$ are both mesoscopic. Here, we vary the microscopic time step $\Delta t_{\M}$ of the HYB model, and we observe the resulting sensitivity. Smaller time steps lead to fewer molecules at the RX (as expected; $n\sqrt{2\Dx{}\Delta t_{\M}}$ decreases but entering the MESO regime introduces the same uncertainty in a molecule's location, leading to a net migration out of $\Vol{1}$). These results highlight the caution that must be taken when using hybrid models. We emphasize that we implemented simple transition rules and that accuracy can be improved by optimizing the transition rules for a given time step as described in \cite{RefWorks:851}.

\begin{figure}[!tb]
	\centering
	\includegraphics[width=\linewidth]{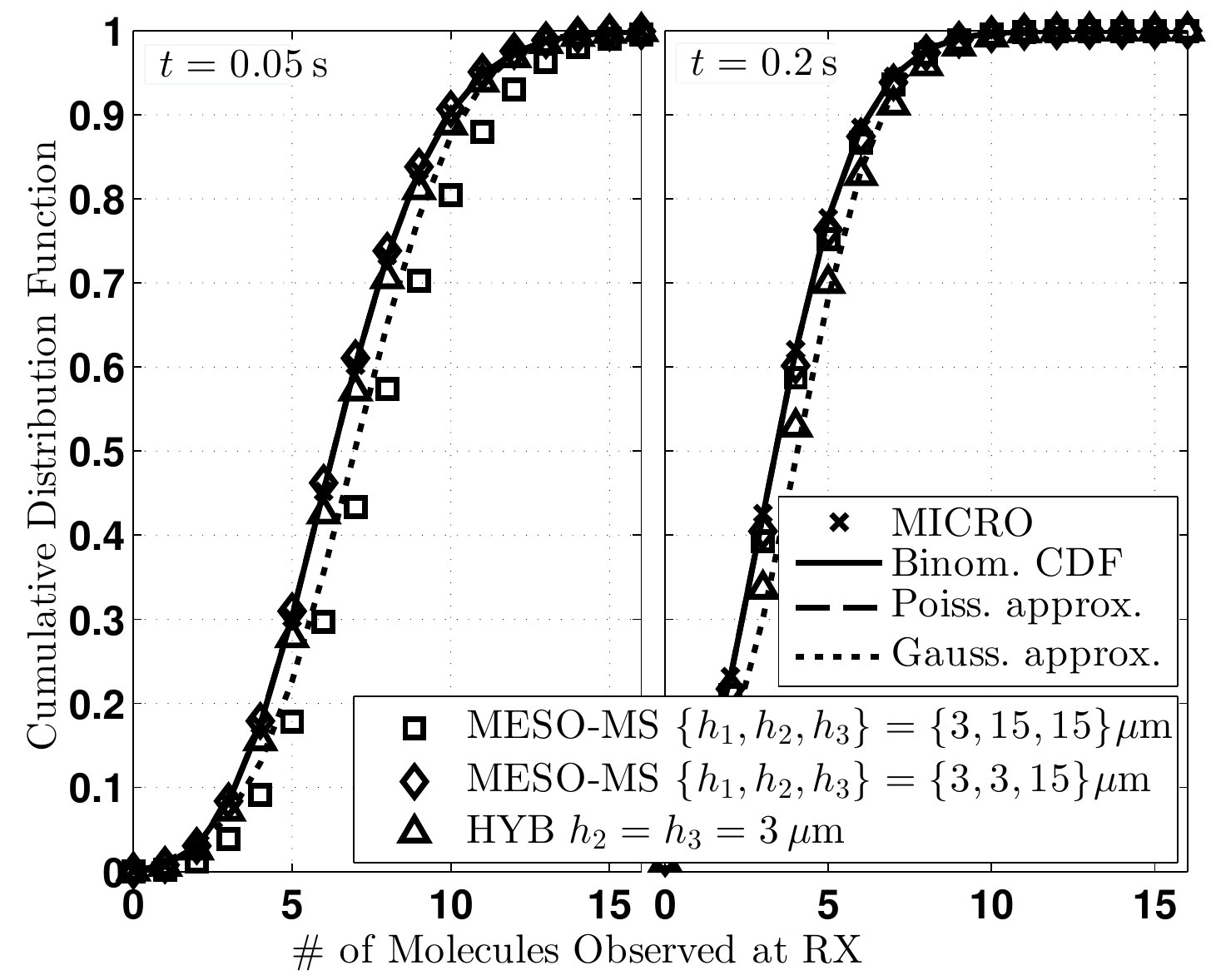}
	\caption{CDF of simulation 3, i.e., the ``point-to-point'' diffusion test.}
	\label{fig_sim_01_p2p_cdf}
\end{figure}

In Fig.~\ref{fig_sim_01_p2p_cdf}, we consider the empirical CDF of the third simulation, evaluated at times $t=0.05\,$s and $t=0.2\,$s after the release by the TX, i.e., near the time of the peak of the expected signal and after the signal is expected to have decreased by more than 3\,dB from the peak value, respectively. At those times, from (\ref{eq_mol_exp_p2p}), we expect 6.98 and 4.05 molecules, respectively. For clarity, we only plot the empirical CDFs for the MICRO model and the least accurate models presented in Fig.~\ref{fig_sim_01_p2p_avg}, since the CDFs for the MESO model with subvolumes of the same size and the HYB model with microscopic $\Vol{2}$ are identical to that of the MICRO model. The MICRO model matches the Binomial CDF at both observation times, and the Poisson approximation is effectively identical to the Binomial CDF. The simulation models that did not accurately capture the expected channel response also did not accurately match the Binomial CDF. Finally, we observe that the Gaussian approximation of the Binomial CDF is almost as poor as the least accurate simulation at both observation times, i.e., the MESO-MS model with $h_2=15\,\mu$m at $t=0.05\,$s and the HYB model that is mesoscopic in $\Vol{2}$ at $t=0.2\,$s.

In the fourth simulation, we consider first-order degradation only and do not allow molecules to diffuse. This test emphasizes the accuracy of simulating chemical reactions alone. We simulate the degradation of 1000 molecules when the reaction constant is $\kth{}=3\,\second^{-1}$. The time-varying response, averaged over $10^4$ realizations, is observed in Fig.~\ref{fig_sim_01_p2p_cdf} for the MICRO models with different values of $\Delta t_{\M}$ and the MESO model. We do not include a curve for the expected response, as given by (\ref{eq_mol_exp_rxn}), but it is identical to the curve shown for the MESO model. We observe that the MICRO model is also very accurate for $\Delta t_{\M}$ varying over orders of magnitude, and the loss of accuracy when $\Delta t_{\M} = 50$ms is only an artifact because this time step is longer than the observation period of $20\,\metre\second$.

\begin{figure}[!tb]
	\centering
	\includegraphics[width=\linewidth]{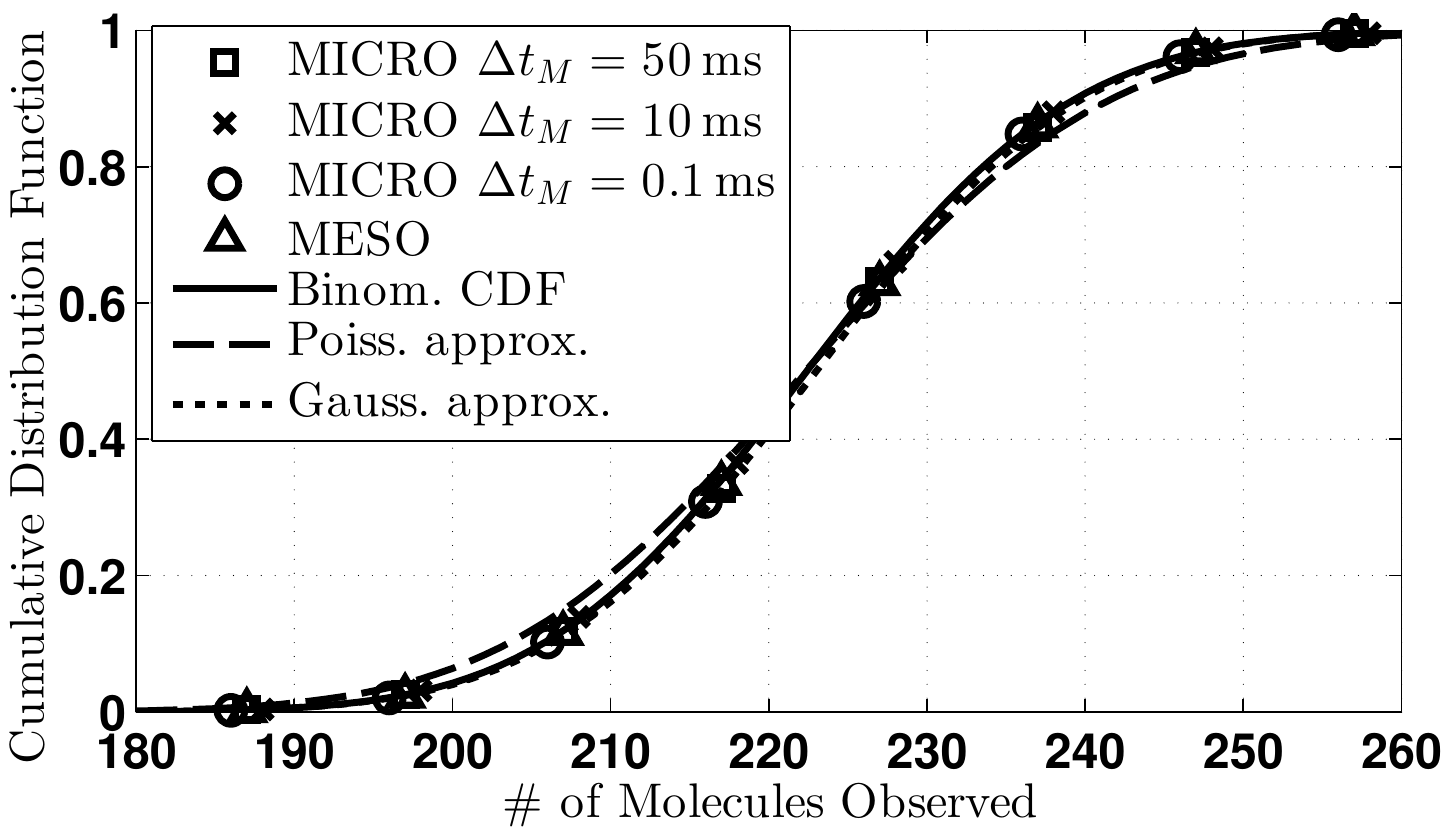}
	\caption{CDF of simulation 4, i.e., the first-order reaction test, where the observation is made at time $t=0.5\,$s. Diffusion is not modeled.}
	\label{fig_sim_06_rxn_cdf}
\end{figure}

In Fig.~\ref{fig_sim_06_rxn_cdf}, we observe the empirical CDF of the fourth simulation for the observation made $t=0.5\,$s after the start. From (\ref{eq_mol_exp_rxn}), about 223 molecules are expected to remain. All of the simulation models agree with the Binomial CDF, and the Gaussian approximation is much more accurate than the Poisson approximation. Here, the underlying trial success probability is 0.223.

\begin{figure}[!tb]
	\centering
	\includegraphics[width=\linewidth]{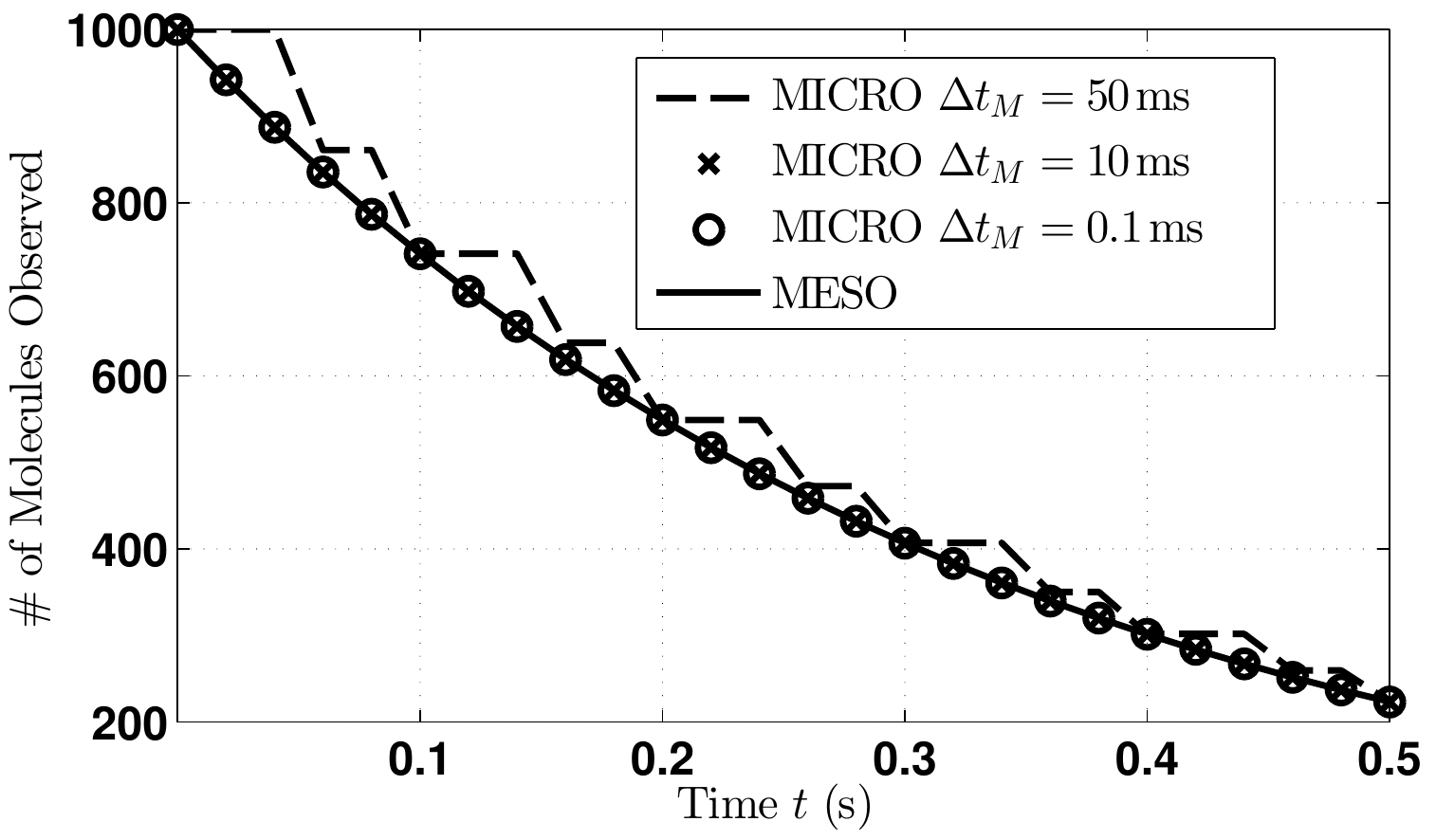}
	\caption{Time-varying channel response of simulation 4, i.e., the first-order reaction test. Diffusion is not modeled.}
	\label{fig_sim_06_rxn_avg}
\end{figure}

\begin{figure}[!tb]
	\centering
	\includegraphics[width=\linewidth]{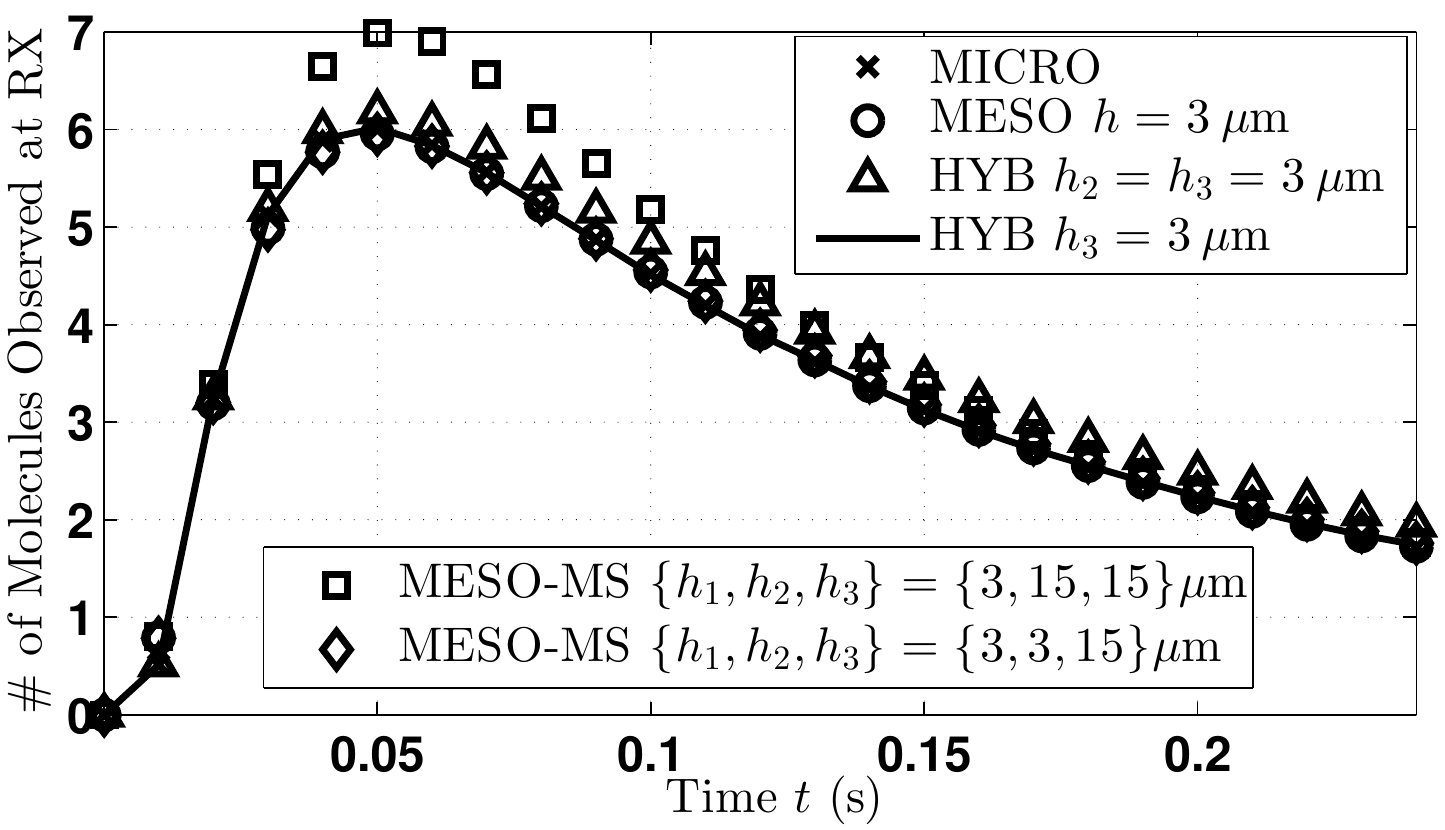}
	\caption{Time-varying channel response of simulation 5, i.e., the ``point-to-point'' reaction-diffusion test.}
	\label{fig_sim_05_p2prxn_avg}
\end{figure}

In the fifth and final simulation, we combine the ``point-to-point'' diffusion test with first-order degradation. 1500 molecules are released at the TX and then observed over time at the RX when the reaction constant is $\kth{}=3\,\second^{-1}$. Simulation results are averaged over $10^5$ realizations. We observe that the accuracy of the simulations is consistent with that observed in the corresponding diffusion-only case. In Fig.~\ref{fig_sim_05_p2prxn_avg}, we observe the time-varying response for the same simulation models that we considered in the ``point-to-point'' diffusion test without degradation. We do not plot the expected time-varying channel response, as given by the product of (\ref{eq_mol_exp_p2p}) and $\EXP{-\kth{}t}$, because it is effectively the same as the average MICRO simulation.
As in Fig.~\ref{fig_sim_01_p2p_avg}, the MESO model, the MESO-MS model with $h_2=3\,\mu$m, and the HYB model where regions $\Vol{1}$ and $\Vol{2}$ are microscopic yield average simulation results that are very similar to the MICRO model. The remaining simulation models are noticeably less accurate than the MICRO model.

\begin{figure}[!tb]
	\centering
	\includegraphics[width=\linewidth]{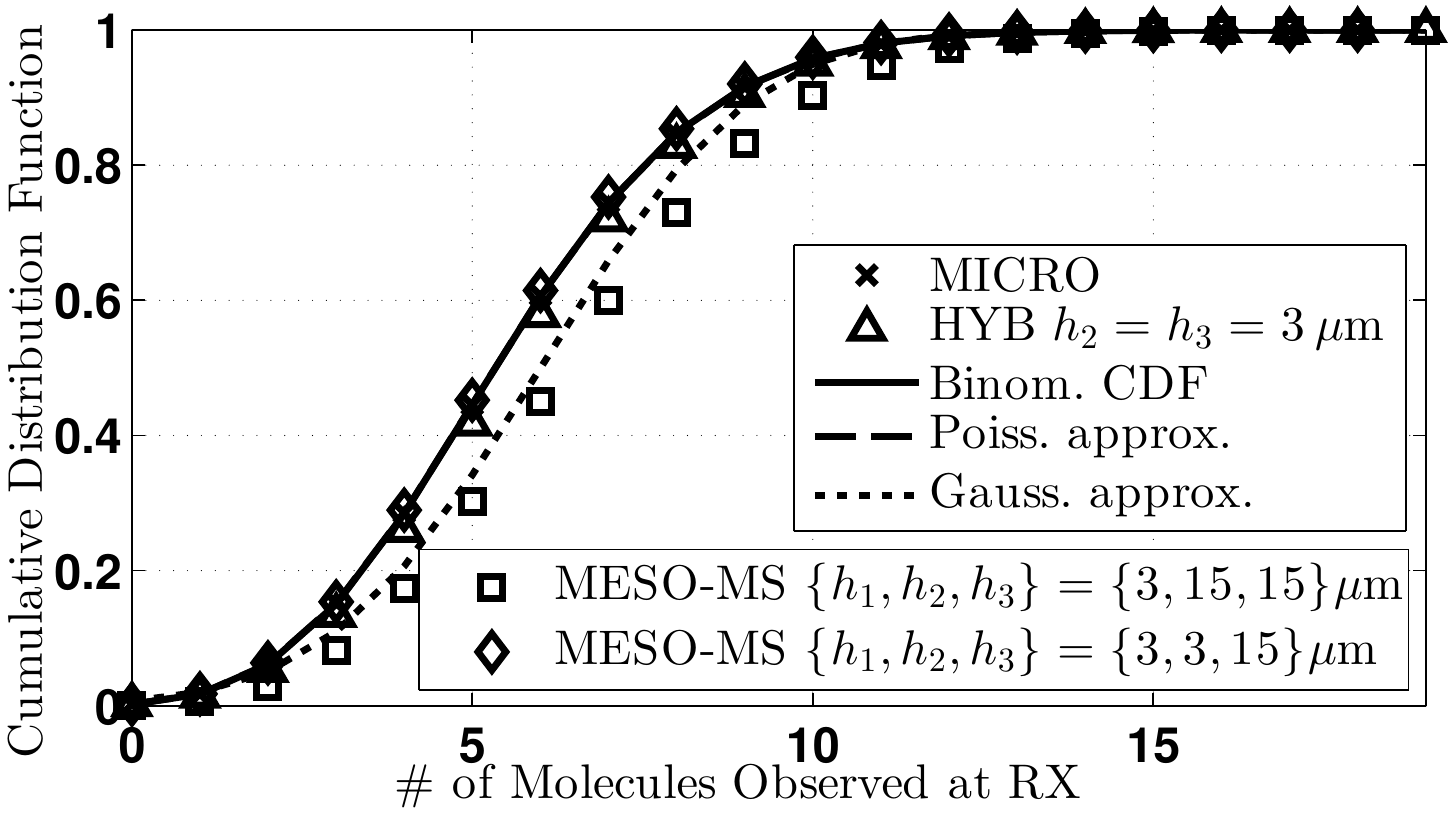}
	\caption{CDF of simulation 5, i.e., the ``point-to-point'' reaction-diffusion test, where the observation is made at time $t=0.05\,$s.}
	\label{fig_sim_05_p2prxn_cdf}
\end{figure}

In Fig.~\ref{fig_sim_05_p2prxn_cdf}, we observe the empirical CDF of the fifth simulation for the observation made $t=0.05\,$s after molecules are released by TX. At that time, from (\ref{eq_mol_exp_p2p}) and $\EXP{-\kth{}t}$ or from inspection of Fig.~\ref{fig_sim_05_p2prxn_avg}, about 6 molecules are expected. As in Fig.~\ref{fig_sim_01_p2p_cdf}, we compare the least accurate simulation models with the MICRO model, whose empirical CDF is once again equivalent to the Binomial CDF. Also, the Poisson approximation of the Binomial CDF is again much more accurate than the Gaussian approximation.

\section{Conclusions}
\label{sec_concl}

In this paper, we compared simulation models to assess their accuracy in diffusion, first order reaction, and first-order reaction-diffusion simulations. We observed the time-varying channel response and the empirical CDF at specific time instants. The microscopic model and the mesoscopic model were generally accurate and their statistics very closely matched those of the underlying Binomial CDF for all simulations. We demonstrated that multi-scale and hybrid models could also maintain accuracy, unless we reduced the computational complexity too close to the communication link. Overall, the statistical accuracy of the receiver was not affected if the hybrid interface or transition to larger subvolumes was as far from the transmitter and receiver as the distance between the transmitter and receiver.

We also compared the suitability of the Poisson and Gaussian approximations of the Binomial CDF, since these approximations are commonly applied in communications analysis. When a large fraction of the released molecules are expected at the receiver, the Gaussian approximation is more accurate. When a small fraction of molecules are expected, the Poisson approximation is more accurate.

Our on-going work is the development of a molecular simulator based on the models presented in this paper and the motivation in \cite{RefWorks:891}. Future implementation includes extension to three dimensions, modeling fluid flow, and implementing more accurate rules for transitions between the microscopic and mesoscopic regimes.

%
\bibliographystyle{abbrv}
\bibliography{../references/nano_ref}  
%
%
\end{document}